\documentclass[12pt, preprint]{aastex}

\slugcomment{Not to appear in Nonlearned J., 45.}

\newcommand{\SI}{[SII]$\lambda$6731}
\newcommand{\Ox}{[OI]$\lambda$6300}

\newcommand{\Ha}{H$\alpha$}
\newcommand{\SII}{[SII]$\lambda\lambda$6716,6731}
\newcommand{\NII}{[NII]$\lambda$6583}
\newcommand{\OI}{[OI]$\lambda\lambda$6300,6363}

\newcommand{\km}{kms$^{-1}$}

\newcommand{\ls}{LS-RCrA 1\ }
\newcommand{\J}{M$_{JUP}$}

\begin{document}


\title{Uncovering the Outflow Driven by the Brown Dwarf LS-RCrA 1: H$\alpha$ as a Tracer of Outflow Activity in Brown Dwarfs.\altaffilmark{1}}


\author{E.T. Whelan\altaffilmark{2}}

\author{T.P. Ray\altaffilmark{2}}

\author{F. Bacciotti\altaffilmark{3}}

\altaffiltext{1}{Based on data collected by UVES observations (observing runs 67.C-0549(B), 69.B-0126(A), 71.C-0429(C) and 
71.C-0429(D)) at the VLT on Cerro Paranal (Chile) which is operated by the European Southern Observatory (ESO).}

\altaffiltext{2}{School of Cosmic Physics, Dublin Institute for Advanced Studies}
\altaffiltext{3}{Osservatorio Astrofisico di Arcetri}


\begin{abstract}

It is now apparent that classical T Tauri-like outflows commonly accompany the formation of young brown dwarfs. To date two optical outflows have been discovered and results presented in this paper increase this number to three. Using spectro-astrometry the origin of the \ls forbidden emission lines in a blue-shifted outflow is confirmed. The non-detection of the red-shifted component of the outflow in forbidden lines, along with evidence for some separation between low and high velocity outflow components, do not support the hypothesis that \ls has an edge-on accretion disk. The key result of this analysis is the discovery of an outflow component to the \Ha\ line. The \Ha\ line profile has blue and red-shifted features in the wings which spectro-astrometry reveals to also originate in the outflow. The discovery that \Ha\  emission in BDs can have a significant contribution from an outflow suggests the use of \Ha\ line widths as a proxy of mass accretion in BDs is not clear-cut. This method assumes that any contribution to the \Ha\ line flux from a possible outflow is negligible. Finally the fact that the \Ha\ line traces both lobes of the outflow while only the blue-shifted lobe is seen in forbidden emission points to the presence of a dust hole in the accretion disk of LS-RCrA 1. This is commonly seen in CTTSs and is assumed to signal the onset of planet formation.

\end{abstract}





\section{Introduction}

The study of brown dwarfs (BDs) in star forming regions has a special importance as 
it not only allows the formation and evolution of BDs to be investigated but also addresses whether the low mass paradigm for star formation can be extended into the substellar regime. Much of what is known about low mass star formation, and in particular the connection between magnetospheric accretion and outflow activity, comes from studying the classical T Tauri stars (CTTS) \citep{Edwards08, Ray07}. Observational studies point to strong similarities between the evolution of young BDs and CTTSs. To summarise BDs are now known to have disks \citep{Wilking99, Natta01, Natta02}, to exhibit T Tauri-like accretion activity and variability \citep{Natta04, Scholz06} and to drive outflows. To date two BD optical outflows \citep{Whelan05, Whelan07} and a two molecular outflows \citep{Phan08, Bourke05} have been reported. Further investigation is needed in order to properly compare BD outflows to protostellar outflows.

\ls  has a spectral type of M6.5 and an estimated mass of 35 - 72 \J\  placing it within the BD mass range \citep{Barrado04, Scholz06}. The 10$\%$ width of the \Ha\ line is commonly used as a test of accretion activity in BDs. BDs with 10$\%$ width greater than 200 \km\ (and equivalent width $>$ 10\AA) are classified as accretors and below as non-accretors \citep{Jay03}. The \Ha\ 10$\%$ width and equivalent width (EW) clearly demonstrate that \ls is an intense accretor \citep{Scholz06, Fernandez01}. Analysis of the spectra discussed here place the \Ha\ 10 $\%$ width between 270\km and 300\km. \cite{Scholz06} give the EW as varying between 44 \AA\ and 125 \AA. Forbidden emission lines (FELs) have proved to be effective tracers of outflow activity in CTTSs and to date they have been used to  explore the kinematics, morphology and physical conditions of jets, at high angular resolution \citep{Ray07}. The so-called traditional tracers of CTT jets, i.e. \OI, \SII, \NII\ are found in the spectra of young BDs. This finding was the first indication that BDs commonly launch outflows. In addition to being an intense accretor \ls was the first BD shown to have FELs and notably it has the strongest forbidden emission of any BD studied to date \citep{Fernandez01, Fernandez05}. 

A further interesting feature of \ls is its sub-luminous nature. \cite{Fernandez01} first identified this object and noted that its luminosity is far less than other objects of similar spectral type. \ls is a strong accretor and current evolutionary models do not take accretion activity into account. \cite{Fernandez01} argue that this could lead to an over-estimation of the age of LS-RCrA 1, thus explaining the lack of agreement between its spectral type and luminosity. Also refer to \cite{Tout99}. An alternative explanation is that \ls could have a near edge-on disk, leading to its photosphere being predominantly seen in scattered light \citep{Barrado04}. As a consequence it would appear fainter (and older) than it really is. 

In this letter the results of a spectro-astrometric analysis of key lines in the spectrum of LS-RCrA 1 are discussed. Spectro-astrometry has been used previously to find outflows driven by the BDs ISO-Oph 102 \citep{Whelan05} and 2MASS1207-3932 \citep{Whelan07}. The molecular component to the ISO-Oph 102 outflow has recently been detected \citep{Phan08}. Although \ls has strong forbidden lines no spatial extension in the line regions has been directly observed \citep{Fernandez05}. Here the displacement of the forbidden emission and \Ha\ line wings from the BD position is mapped, confirming their origin in an outflow driven by LS-RCrA 1. Results contribute to the debate over the cause of the sub-luminous nature of \ls and increase the number of confirmed BD optical outflows to three.

\section{Observations and Analysis}

The high resolution (R=57,000) spectra presented here were obtained with the UV-Visual Echelle Spectrometer (UVES) 
(Dekker et al. 2000) on the ESO VLT UT2 and first published by \cite{Fernandez05}. For the purpose of spectro-astrometric analysis the raw data were obtained from the ESO Archive. Observations of \ls were taken over a range of dates between June 3$^{rd}$ 2003 and July 4$^{th}$ 2003 and at a range of slit position angles (P.As). As reported by \cite{Fernandez05} the range of P.A.s resulted from the setting of the parallactic angle so as to minimize losses due to atmospheric differential refraction (refer to Table \ref{tab1}). The seeing over all nights varied between 0\farcs9 and 1\farcs7 and the spatial sampling was 0\farcs182 per pixel. The data were analysed using spectro-astrometry as outlined in \cite{Whelan05} and \cite{Whelan08}.

Spectro-astrometry was first applied to the optical/near-infrared spectra of Herbig Ae/Be and CTTSs \citep{Bailey98, Whelan04}. Application of the technique to the study of low and intermediate mass stars is relatively straight-forward as both the continuum and line emission is strong. However, in the case of BDs the continuum emission in the range 6000-7000 \AA\ is inherently faint and the FELs are far weaker than those excited in CTT jets. In order to increase the signal to noise (S/N) of the BD spectra and thus the spectro-astrometric accuracy, the spectra are summed or smoothed (in the dispersion direction) before the transverse spatial profiles are fitted \citep{Whelan07}. Also as explained in \cite{Whelan05} and \cite{Whelan08} the continuum is fitted and then removed before the offset in the line region is measured. In our study of BDs we sum/smooth the non-continuum subtracted and subtracted spectra separately so that the accuracy to which we measure the position of the line and continuum emission is comparable. The method chosen to increase the S/N does not effect the overall results. In Figures 1 and 2 each point is the centroid of a spatial profile extracted at a particular velocity and each extracted profile is the sum of a number of adjacent profiles. Hence offsets are a moving sum across the continuum and continuum subtracted line region. 

In the case of LS-RCrA 1, while a high S/N was achieved for the forbidden emission and \Ha\ lines the continuum emission was relatively very faint (recall that \ls is described as sub-luminous). Indeed, as reported in \cite{Whelan06}, spectra of \ls obtained using the Magellan Inamori Kyocera Echelle (MIKE) on the Magellan II Telescope were found to be unsuitable for spectro-astrometric analysis, simply due to the non-detection of the continuum emission. Even with the better S/N UVES spectra obtained by \cite{Fernandez05} and increasing the S/N (as described above), it was still only possible to recover the continuum position in the region of the \Ha, \SII\ and [NII]$\lambda$6583 lines but not at bluer wavelengths.

Also note that as in previous investigations of BD outflows with UVES, spectro-astrometric artifacts \citep{Brannigan06} were not found to be a problem. In \cite{Whelan05} and \cite{Whelan07} artifacts were ruled out through the demonstration that lines like HeI $\lambda$6678 were not offset. If measured offsets were indeed artifacts then displacements would be found in all lines, even those which are chromospheric in origin. In this study spectra are analysed at several PAs and overall measured offsets in the lines are on a scale of 50-200 mas indicating the presence of spatially extended emission originating in an outflow. Offsets vary with position angle as expected, providing confirmation that they are real.

 \begin{table}
\begin{tabular}{|clcc|}       
 \hline\hline 
 PA (avg)                             &Date (2003)   &Line &Maximum Offset (mas)      
 \\ 
\hline
45$^{\circ}$         &25-06                 &[NII]$\lambda$6548, [SII]$\lambda$6731                       &150 $\pm$ 14, 140  $\pm$  9
\\
                           &                 &\Ha\                                               &60, -45   ($\pm$ 14)                  
 \\
 \hline
63$^{\circ}$            &3-06             &[NII]$\lambda$6548, [SII]$\lambda$6731   &100 $\pm$ 25, 110  $\pm$ 12      
\\ 
   &                 &\Ha\                                           &50, -28 ($\pm$ 7)
\\ 
 \hline
74$^{\circ}$            &09-06             &[NII]$\lambda$6548, [SII]$\lambda$6731   &100 $\pm$ 24, 100  $\pm$ 12       
\\ 
   &                 &\Ha\                                               &15, -15 ($\pm$ 5)
\\
 \hline
82$^{\circ}$            &13-06             &[NII]$\lambda$6548, [SII]$\lambda$6731   &40 $\pm$ 5, 45  $\pm$  9         
\\ 

   &                 &\Ha\                                               &10, -10 ($\pm$ 4)
\\
 \hline
-73$^{\circ}$            &04-07             &[NII]$\lambda$6548, [SII]$\lambda$6731   &-80 $\pm$ 19, -80 $\pm$ 10      
\\ 
   &                 &\Ha\                                              &-5, 5 ($\pm$ 6)
\\
 \hline  
\end{tabular}
\caption{Spatial Offsets measured in the blue-shifted \NII\ and \SI\ lines and the blue/red \Ha\ line wings at slit PAs between 45$^{\circ}$ and -73$^{\circ}$.}
\label{tab1}

\end{table}

\section{Results and Discussion}
\label{sectionR}

Table \ref{tab1} lists the maximum spectro-astrometric offsets in the blue-shifted [NII]$\lambda$6583 and \SI\ lines and the blue/red \Ha\ wings, for the range of slit PAs used. In addition, the spectro-astrometric analysis at 45$^{\circ}$ is presented in Figures 1 and 2. Offsets are with respect to the continuum positon with positive offsets to the North and negative to the South. \cite{Fernandez05} give the average velocities of the \ls FELs at $\sim$ -5 \km, $\sim$ -22 \km\ and $\sim$ -13 \km\ for the [OI]$\lambda$6300, [NII]$\lambda$6583 and [SII]$\lambda$6731 lines respectively. Our reduction of their data yields the same average values.  Uniquely the outflow is also detected in \Ha.  Blue and red-shifted ``humps" are observed at velocities of $\sim$ -50 \km\ and + 100 kms$^{-1}$ and these are found to be offset in the direction of the blue-shifted outflow and red-shifted counter-flow respectively, as defined by the FELs (refer to Table \ref{tab1} and Figure 2). It is clear that the PA of 45$^{\circ}$ lies closest to the actual PA of the \ls outflow as offsets are maximised at this slit position. The change in displacement with PA allows the PA of the outflow to be constrained.  As the slit is moved from 45$^{\circ}$ to 82$^{\circ}$ displacement decreases from $\sim$ 145 mas to 45 mas in the FELs and $\sim$ 50 mas to 10 mas in \Ha. From this the offset one would expect to measure at P.As of 90$^{\circ}$ and 0$^{\circ}$ is calculated.  These two values place the outflow PA at $\sim$ 15$^{\circ}$.

Note that while only the blue-shifted outflow is traced by the FELs both the blue and red-shifted lobes are traced by \Ha. A similar trend has been seen for some CTTSs and it is hypothesised that while the circumstellar disk obscures the red-shifted forbidden emission, the permitted emission (which originates much closer to the star due to its significantly higher critical density) is visible through a hole or gap in the disk \citep{Takami01, Whelan04}. We estimate that a minimum (projected) disk radius of 0\farcs15 ($\leq$ 22.5 AU at the distance of the Coronae Australis star forming region) is needed in order to hide the redshifted component to the \ls forbidden emission.  The formation of disk gaps in CTTSs is believed to result from grain growth in the inner disk or clearing due to an orbiting planetesimal \citep{Whelan04}. For CTTSs and young BDs grain growth is probably the most likely explanation and signals the onset of planet formation \citep{Apai05}. Previous studies have argued for the occurrence of planet forming processes in BD disks \citep{Muz06, Scholz07}. The discovery of a dust gap in the disk of LS-RCrA 1, large enough for the red-shifted outflow to be detected, supports the hypothesis that BD disks could at some stage harbour planets. 

Spectro-astrometric studies of CTTSs and Herbig Ae/Be stars with disk dust holes have allowed the radii of the holes to be estimated \citep{Takami01, Whelan04}. The measured red-shifted offset is taken as an estimate of the radius of the dust hole. In all cases estimates of hole radius from spectro-astrometry were compared with estimates made from the spectral energy distributions (SEDs) of the stars. The SEDs of stars with circumstellar disks show an excess of emission above the stellar emission (originating from the disk). Stars with inner disk holes, have dips in this excess emission, corresponding to the distance from the star that the region free from dust extends to. Both methods of measuring the radii of the dust holes have yielded similar results \citep{Takami03}. For \ls a red-shifted offset of 45 mas $\pm$ 14 mas at 45$^{\circ}$ indicates a gap of $\sim$ 6.75 $\pm$ 2 AU in radius. This is somewhat larger than the gap observed by \cite{Muz06} in the BD L316 (this estimate was made from the SED of the BD).  
Analysis of the SED of \ls and of spectra taken at the outflow PA is needed in order to constrain the size of its inner dust hole.

\cite{Barrado04} argue that \ls has an edge-on disk, explaining its faintness compared to other objects of the same spectral type. For an outflow source with an edge-on disk and thus an outflow in the plane of the sky one would expect to see both lobes of the outflow and to measure low radial velocities for any lines tracing the outflow. The high and low velocity components (HVC, LVC) to FELs, commonly detected in the spectra of CTTSs \citep{Hirth97}, appear blended for high jet inclination angles. A good example is the outflow driven by the BD 2MASS1207-3932. This BD is known to have a near edge-on disk and the analysis of its outflow supported this. \cite{Whelan07} detected blue and red-shifted [OI]$\lambda$6300 emission at velocites of +8\km\ and -4 \km. The lack of red-shifted forbidden emission means that \ls does not have an edge-on disk but that the disk is sufficiently inclined to obscure the red lobe. 

There is also some evidence of separation between the low and high velocity forbidden emission. The differing origins of the LVC and HVC is well documented for CTTSs and results in [NII]$\lambda$6583 only tracing the HVC (well collimated high velocity jet) and both [SII]$\lambda$6731 and \Ox\ tracing both the LVC (less collimated low velocity wind) and HVC \citep{Hartigan95}. Although \Ox\ due to its much higher critical density is the better tracer of high density regions close to the star and hence the LVC. The average radial velocities listed above for the \Ox, [NII]$\lambda$6583 and [SII]$\lambda$6731 lines demonstrate that the \Ox\ line primarily traces the LVC (-5 \km), the [NII]$\lambda$6583 line the HVC only (-22 \km\ and wing extended to -75\km) and the \SI\ line the LVC and HVC (peak at -13 \km\ and wing extended to -40 \km). This is supported by the spectro-astrometric analysis of the \SI\ and [NII]$\lambda$6583 lines at $\sim$ 45$^{\circ}$. The maximum offset in the [NII]$\lambda$6583 line occurs at the peak velocity of $\sim$ -22 \km, however, for the [SII]$\lambda$6731 the maximum offset is in the wing. This points to the fact that the [SII] emission making up the wing originates in the high velocity component of the outflow, while the bulk of the emission traces a slower less extended component.

The \ls FELs are reminiscent of the FELs lines found in the spectrum of the CTTS DG Tau. The DG Tau \Ox\ and \SI\ lines have a low velocity primary peak and a high velocity secondary peak with maximum offsets occurring in the high velocity peaks. The \Ox\ line peaks occur at $\sim$ -47 \km\ and -266 \km\ and the \SI\ line peaks at $\sim$ -47 \km\ and -240 \km. The \NII\ line is single peaked and maximum offsets occur at the peak velocity of -240 \km.  The spectro-astrometric analysis of the DG Tau \Ox\ and \SI\ lines is published in \cite{Whelan04}. The analysis of the \NII\ was not published but values quoted here come from the same data. Overall the kinematical and spectro-astrometric results do not support the argument that \ls has an edge-on disk. As a result the questions related to the luminosity of this source persist. The fact that it is a strong accretor is now the most likely explanation of the lack of agreement between its spectral type and luminosity.

Finally, the spectro-astrometric analysis of the \Ha\ line shall be discussed in the context of the use of \Ha\ line widths as a measure of mass accretion in BDs.  In addition to being used as a diagnositc of on-going accretion, model fitting of the \Ha\ line is used to estimate the rate of mass accretion \citep{Muz03}. The detection of a significant outflow component to the \ls \Ha\ line cautions against the use of \Ha\ alone as a measure of mass accretion rates. Other methods of estimating the mass accretion rate are based on measurements of veiling and of the flux of the CaII$\lambda$8662 line \citep{Mohanty05}. It is clearly important to rule out any significant contribution to the \Ha\ line from a jet/outflow before using it as a reliable probe of magnetospheric accretion.

\section{Summary}

\ls was the first young BD found to exhibit signatures of outflow activity in the form of FELs, similar to those known to trace classical T Tauri jets. Although the \OI, \SII\ and [NII]$\lambda$6583 lines are strong, spectro-scopic and imaging studies failed to find any extension in these line regions in the form of an outflow \citep{Fernandez05}. Furthermore LS-RCrA 1 which is noted to be faint in comparison to other sources of the same spectral type, was shown to be a remarkably strong accretor. Several papers have attempted to reconcile the strong outflow and infall signatures with the sub-luminous nature of this object and have offered an edge-on disk scenario as the most likely explanation. 

The spectro-astrometric analysis presented here greatly adds to our understanding of this exemplar object as follows. Firstly, a spatial offset in the \SI\ and [NII]$\lambda$6583 FEL regions, from the continuum, is seen, verifying their excitation in a blue-shifted outflow and constraining the P.A. of the outflow at $\sim$ 15$^{\circ}$. The velocities and offsets measured in the lines at a P.A. of 45$^{\circ}$ (closest to the outflow P.A.) show the lines to be tracing different components of the overall outflow. The [NII]$\lambda$6583 is clearly tracing the higher velocity component (associated with a jet in CTTSs) while the \SI\ line is a mixture of high and low velocity emission and the maximum offsets occur in the blue-shifted line wing. The complete lack of red-shifted forbidden emission along with some evidence of separation between low and high velocity emission means that \ls cannot have an edge-on disk. 

Secondly it is demonstrated for the first time that \Ha\ can trace outflows driven by BDs and that the contribution to the line can be significant. This discovery is of consequence as \Ha\ line widths are frequently used as an indicator of accretion activity (10$\%$ width) and a measure of mass accretion rate in BDs. Any study using \Ha\ to probe mass accretion must consider the possibility of a significant outflow component to the line. Spectro-astrometry is the best tool to use to disentangle the outflow component to any permitted emission line. Finally evidence of a size-able dust-hole in the disk of \ls exists supporting the hypothesis that BD disks could possibly harbour planets.

Overall, results support the continuation of key processes in the formation and evolution of low mass stars into the substellar regime. Again it is clear that outflows accompany the formation of BDs and that they are analogous but scaled-down from T Tauri outflows. The established tracers of CTT jets, the optical FELs, probe BD outflows in an equivalent way and can, as demonstrated here, have low and high velocity components. In addition, the processes leading to the formation of debris disks also appear to occur in the disks of BDs.

\begin{center}
\begin{figure}
\includegraphics[width=9cm]{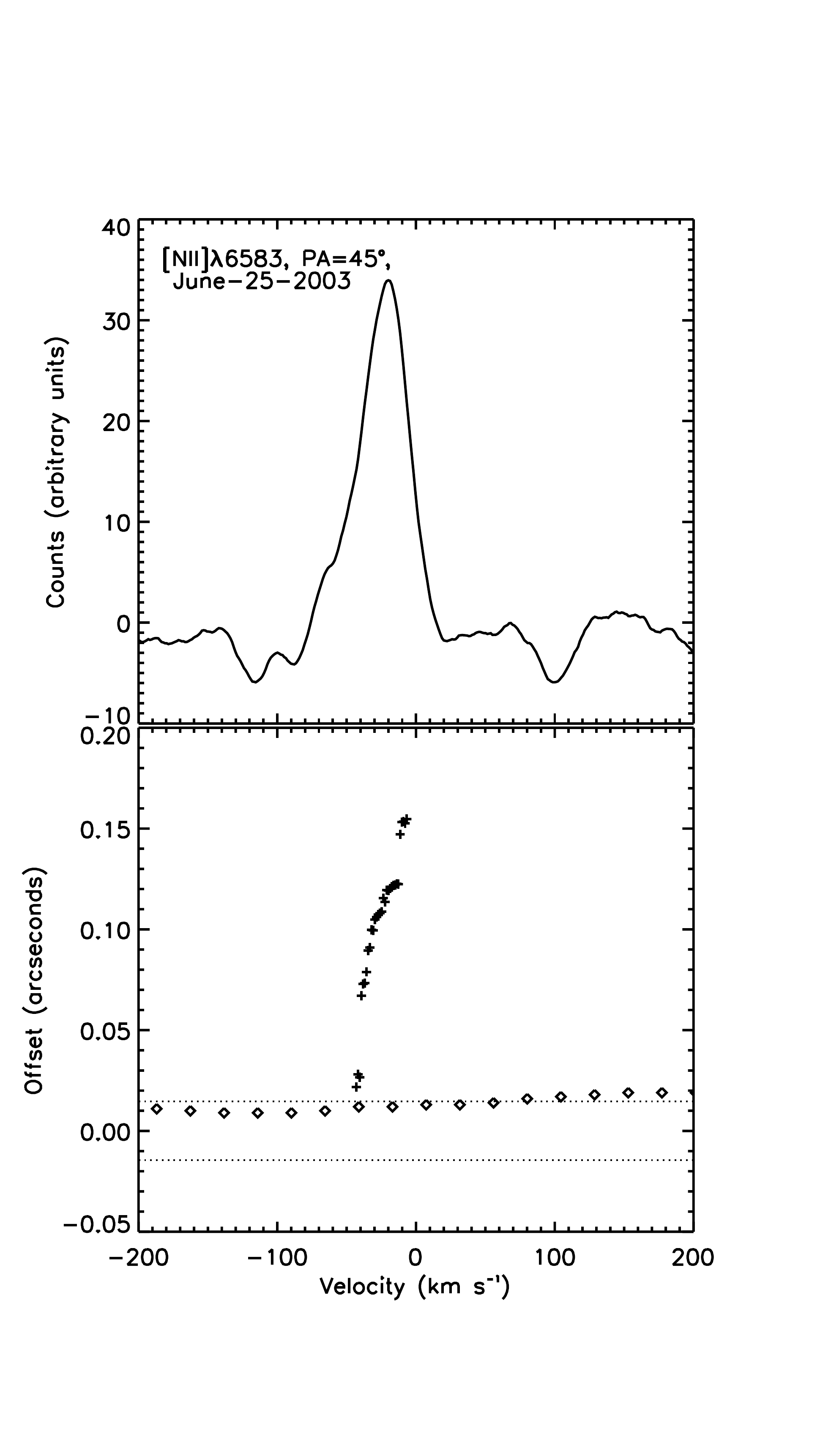}
\includegraphics[width=9cm]{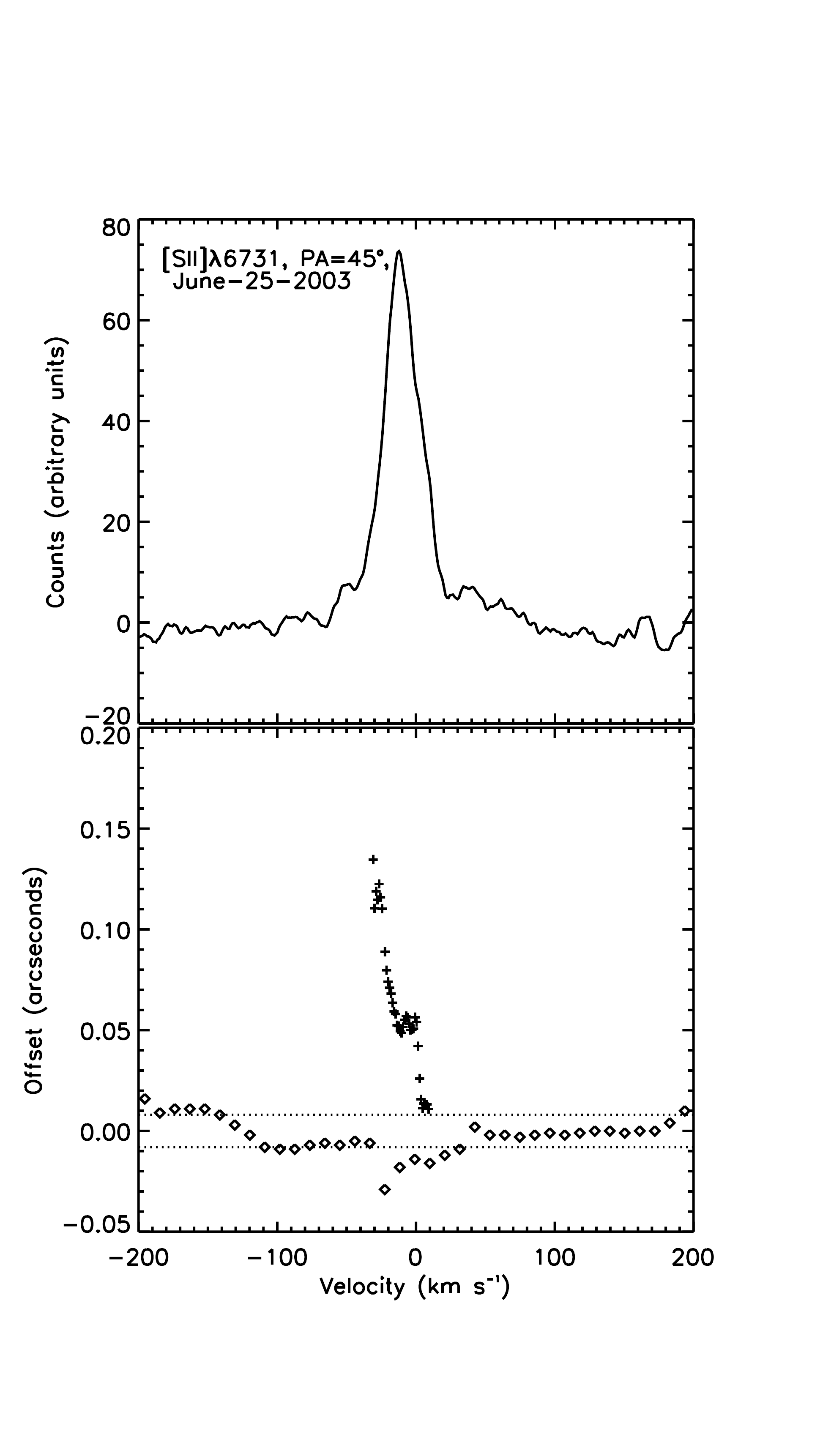}
 \caption{Spectro-astrometry of the blue-shifted \NII\ and \SI\ lines at a slit P.A. of 45$^{\circ}$. As explained the spectra are summed in order to increase the S/N and thus the spectro-astrometric accuracy. 
 The diamonds represent the continuum position measured over a range of velocities and the crosses the offset in the pure line region. All velocities are systemic. The accuracy in the measurement of both the line and continuum position is comparable and the dashed lines de-lineate the $\pm$ 1$\sigma$ error. } 
\end{figure}
 \end{center}

  \begin{center}
 \begin{figure}
\includegraphics[width=9cm]{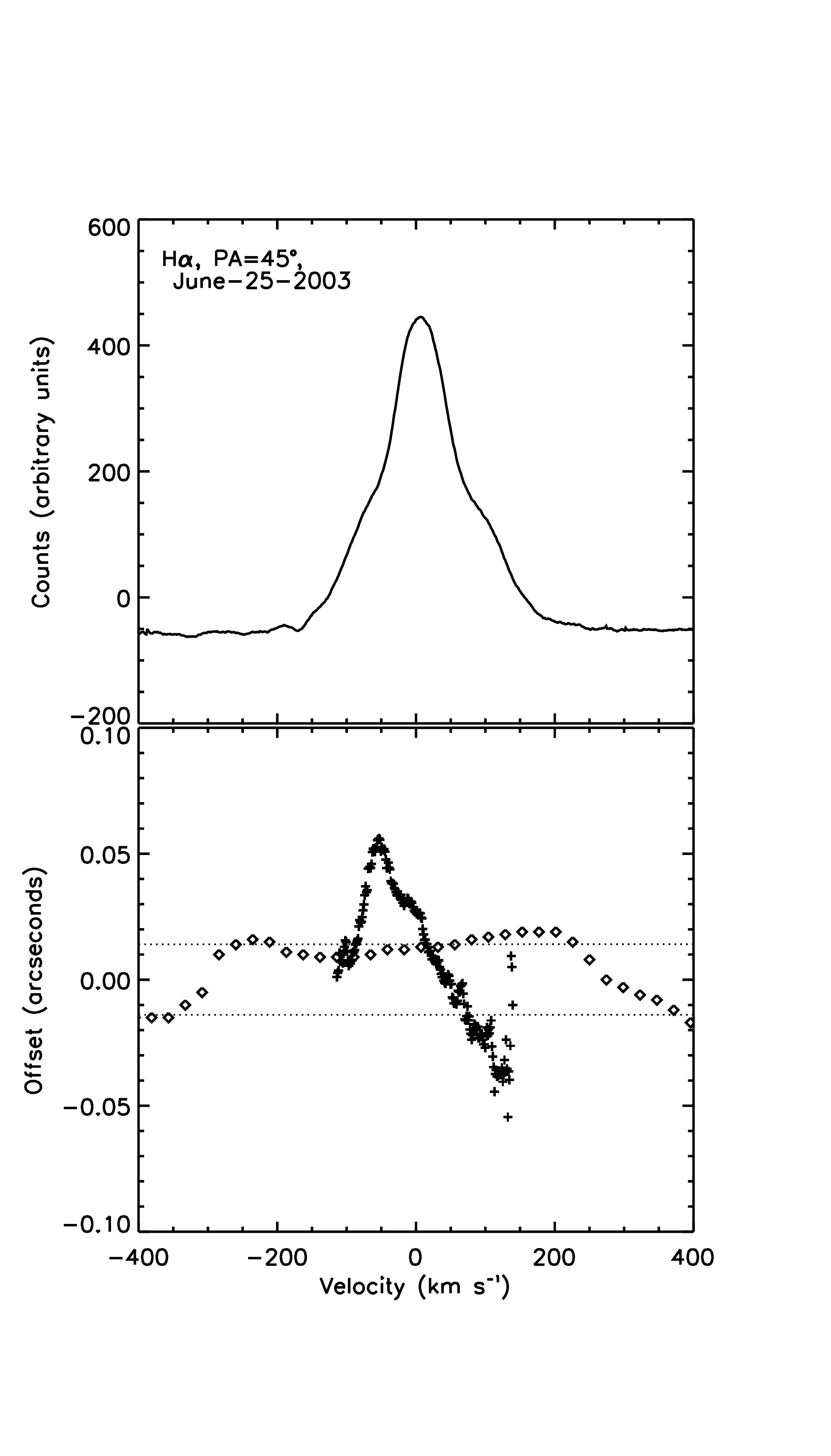}
 \caption{Spectro-astrometry of the \Ha\ line at a slit PA of 45$^{\circ}$. The spectro-astrometric analysis was carried out in the same way as for the FELs. The \ls outflow is clearly contributing significantly to the overall line flux and offsets in the wings are revealed at velocities of $\sim$ -50 \km\ and +100 \km.}
 \end{figure}
 \end{center}

\acknowledgements{The present work was supported in part by the European Community's Marie Curie Actions - Human Resource and Mobility within the 
JETSET (Jet Simulations, Experiments and Theory) network under contract MRTN-CT-2004 005592 and by Science Foundation Ireland (contract no. 04/BRG/P02741). 
This work is based on observations made with the European Southern Observatory
telescopes obtained from the ESO/ST-ECF Science Archive Facility.

}

\end{document}